\documentclass[preprint,preprintnumbers,amsmath,amssymb, nofootinbib]{revtex4}

\usepackage{graphicx}% Include figure files
\usepackage{dcolumn}% Align table columns on decimal point
\usepackage{bm}% bold math
\usepackage{epsfig}
\usepackage{hyperref}
\usepackage{amsmath}
\usepackage{amssymb}

\def\fun#1#2{\lower3.6pt\vbox{\baselineskip0pt\lineskip.9pt
  \ialign{$\mathsurround=0pt#1\hfil##\hfil$\crcr#2\crcr\sim\crcr}}}

\input epsf

\newcommand{\be}{\begin{equation}}
\newcommand{\ee}{\end{equation}}
\newcommand{\bea}{\begin{eqnarray}}
\newcommand{\eea}{\end{eqnarray}}

\newcommand{\ch}[1]{\,{\rm cosh}\,#1}
\newcommand{\sh}[1]{\,{\rm sinh}\,#1}

\def\pslash{\not{\hbox{\kern-2pt p}}}
\newcommand{\mb}{\mathbf}

\begin{document}
\preprint{\tighten \vbox{\hbox{ANL-HEP-PR-10-14} 
                \hbox{FERMILAB-PUB-10-098-T}
                \hbox{NUHEP-TH/10-05}  }}

\vspace*{2cm}

\title{Revealing the electroweak properties of a new scalar resonance}

\vspace*{0.2cm}

\author{
\vspace{0.5cm} 
Ian Low$^{a,b}$ and  Joseph Lykken$^{c}$ }
%\email{ilow@northwestern.edu}
\affiliation{
\vspace*{.2cm}
$^a$ \mbox{High Energy Physics Division, Argonne National Laboratory, Argonne, IL 60439}\\
$^b$ \mbox{Department of Physics and Astronomy, Northwestern University, Evanston, IL 60208} \\
$^c$  \mbox{Fermi National Accelerator Laboratory, P.O. Box 500, Batavia, IL 60510}
\vspace*{0.8cm}}

\begin{abstract}
\vspace*{0.5cm}
One or more new heavy resonances may be discovered in experiments at the CERN Large
Hadron Collider. In order to determine if such a resonance is the long-awaited Higgs boson,
it is essential to pin down its spin, $CP$, and electroweak quantum numbers. Here we 
describe how to determine what role a newly-discovered
neutral $CP$-even scalar  plays in electroweak symmetry breaking,
by measuring its relative decay rates into
pairs of electroweak vector bosons: $W^+W^-$, $ZZ$, $\gamma\gamma$, and $Z\gamma$.
With the data-driven assumption that electroweak symmetry breaking
respects a remnant custodial symmetry, we perform a general analysis with
operators up to dimension five. Remarkably, only three pure cases and one nontrivial mixed
case need to be disambiguated, which can always be done if 
all four decay modes to electroweak vector bosons can be observed or constrained.
We exhibit interesting special cases of Higgs look-alikes with nonstandard
decay patterns, including a very suppressed branching to $W^+W^-$ or very
enhanced branchings to $\gamma\gamma$ and $Z\gamma$. Even if two
vector boson branching fractions conform to Standard Model expectations for
a Higgs doublet, measurements of the other two decay modes could unmask a Higgs
imposter.

\end{abstract}

%\pacs{draft}

\maketitle

\section{Introduction}
Experiments at the Fermilab Tevatron and the CERN Large Hadron Collider are engaged in searches
for the Higgs boson, a heavy scalar resonance predicted by
the Standard Model (SM). SM Higgs bosons are excitations of the neutral $CP$-even component
of an $SU(2)_L$ weak isospin doublet field $H$ carrying unit hypercharge under $U(1)_Y$, whose
vacuum expectation value (VEV) $v/\sqrt{2}$ is responsible for electroweak symmetry breaking 
(for reviews, see \cite{Gunion:1989we,Djouadi:2005gi}).

If one or more new heavy resonances are discovered at the LHC, it will be imperative
to pin down their quantum numbers relative to the expected properties of the SM Higgs.
Determination of the spin and $CP$ properties of a new resonance will be 
challenging, although recent studies indicate that definitive results could be obtained
at or around the moment of discovery, if the decay mode to $ZZ$ is observable \cite{Cao:2009ah, DeRujula:2010ys,Gao:2010qx}.

Given a neutral $CP$-even spin 0 resonance $S$, one still needs to establish its 
electroweak quantum numbers in order to reveal any possible connection to electroweak
symmetry breaking. This in turn requires information about the couplings between
$S$ and pairs of vector bosons, which can be extracted from observations of $S$
decaying via $W^+W^-$, $ZZ$, $\gamma\gamma$, or $Z\gamma$.
To an excellent approximation
the couplings of the SM Higgs boson to $WW$ and $ZZ$ derive from the dimension-four Higgs kinetic terms
in the SM effective action, and are thus directly related to both the strength of electroweak
symmetry breaking and the electroweak quantum numbers of the Higgs field.
The couplings of the SM Higgs boson to $\gamma\gamma$, $Z\gamma$, or a pair of gluons
are elegantly derived from the observation that Higgs couplings in the SM 
are identical to those of a conformal-compensating dilaton in a theory where
scale invariance is violated by the trace anomaly \cite{Adler:1976zt, Djouadi:1993ji, Goldberger:2007zk, Bai:2009ms}.
Thus these couplings
appear first at dimension five, with coefficients related to SM gauge group beta functions.

In this paper we exhibit a general analysis, up to operators of dimension five, of
the relation between the electroweak properties of $S$ and its decay branchings
to $V_1V_2 = WW$, $ZZ$, $\gamma\gamma$, and $Z\gamma$. We ignore decays into two
gluons because of the folklore that these are unobservable, and postpone until
the end a discussion of extracting complementary information from 
vector boson fusion production of $S$ \cite{Plehn:2001nj, Hankele:2006ma}. Nevertheless, we should emphasize that our analysis only involves the decays of the scalar into electroweak vector bosons, and hence is independent of the production mechanism of the scalar. 

A key feature of our analysis is the classification of Higgs look-alikes according
to the custodial symmetry $SU(2)_C$. In the SM this global symmetry is the diagonal remnant
after electroweak symmetry breaking of an accidental global $SU(2)_L\times SU(2)_R$,
in which $SU(2)_L$ and the $U(1)_Y$ subgroup of $SU(2)_R$ are gauged. 
Custodial symmetry implies $\rho\equiv m_W^2/(m_Zc_w)^2 = 1$ \cite{Sikivie:1980hm}, where $c_w$ is
the cosine of the weak mixing angle. Experimentally $\rho$ is constrained to be
very close to one \cite{Amsler:2008zzb}, implying either that the full scalar sector
respects $SU(2)_C$, or that there are percent-level cancellations unmotivated by
symmetry arguments. In our analysis we will assume that unbroken
$SU(2)_C$ is built into the scalar sector.

We consider $S$ arising from one of the neutral $CP$-even components of arbitrary 
spin 0 multiplets of $SU(2)_L\times SU(2)_R$. The case of a singlet under
$SU(2)_L\times SU(2)_R$ is special, since then no $SV_1V_2$ couplings can
appear from operators of dimension four. All other cases can be grouped
according to whether the neutral scalar components transform as a singlet
or a 5-plet under $SU(2)_C$. Again under the assumption that the full scalar
potential respects the custodial symmetry, these three ``pure cases'' only give
rise to one nontrivial mixed case, i.e. when $S$ from a $SU(2)_L\times SU(2)_R$
singlet mixes with another $S$ from the $SU(2)_C$ singlet part of a
$SU(2)_L\times SU(2)_R$ nonsinglet.

Given the framework just described, we are able to enumerate all possible
deviations from the SM expectations for decays of a Higgs look-alike into
pairs of electroweak bosons. These deviations are typically quite large,
and thus accessible to experiment at the LHC. Furthermore
the deviations exhibit patterns that point towards
particular non-SM scenarios. It would therefore be
possible with LHC data to rule out a new scalar resonance as the agent (or the
sole agent) of electroweak symmetry breaking.
This possibility emphasizes the importance of observing all
four $V_1V_2$ decay channels at the LHC with maximum sensitivity.
We give examples of Higgs imposters that meet SM expectations
for branching fractions into two of the electroweak $V_1V_2$ modes,
only revealing their ersatz nature in the other two $V_1V_2$ decay modes. The approach taken here is complimentary to that  in Refs.~\cite{Cao:2009ah, DeRujula:2010ys,Gao:2010qx}, where angular correlations and total decay width were used to distinguish Higgs look-alikes. A fully global analysis using all of the available decay and production observables in each channel will of course give superior results to the simple counting experiments described here.

%\footnote{As pointed out in Ref.~\cite{Cao:2009ah}, the total width of the scalar resonance could be useful as well.}

In Sect. \ref{sect:section2} we describe the dimension four couplings
of an arbitrary  neutral $CP$-even scalar charged under $SU(2)_L\times U(1)_Y$
to $WW$ or $ZZ$; we also describe the general dimension five couplings
of a $SU(2)_L\times U(1)_Y$ singlet to two electroweak vector bosons.
Sect. \ref{sect:section3} contains the general framework based on custodial
symmetry. In Sect. \ref{sect:section4} we provide general results on the patterns
of $S\to V_1V_2$ branching fractions, as well as discussing some interesting special cases.
Further discussion and outlook are in Sect. \ref{sect:section5}, with some general formulae
for off-shell decays relegated to an appendix.

\section{Scalar Couplings with $V_1V_2$}
\label{sect:section2}

In this section we consider scalar couplings with two electroweak gauge bosons $V_1V_2$, where $V_1V_2=\{WW, ZZ, Z\gamma, \gamma\gamma\}$. Such couplings are dictated by the electroweak quantum numbers of the scalar $S$. We will write down $SU(2)_L\times U(1)_Y$ invariant operators giving rise to the $SV_1V_2$ couplings at the leading order.  For an electroweak nonsinglet, the leading operator is the kinetic term of the scalar, assuming $S$ receives a VEV, while for the singlet scalar the leading operator starts at dimension five.

For nonsinglet scalars, the leading contribution to  the $SV_1V_2$ coupling arises from spontaneous breaking of $SU(2)_L\times U(1)_Y$ down to $U(1)_{em}$ via the Higgs mechanism, when $S$ develops a VEV. It is possible to derive the general coupling  when there are multiple scalars in arbitrary representations of the $SU(2)_L$ group \cite{Tsao:1980em, Haber:1999zh}.  Using the notation $\phi_k$ for scalars in the complex representations and $\eta_i$ for scalars in the real representations\footnote{A real representation is defined as a real multiplet with integer weak isospin and $Y=0$. }, the kinetic terms are
\be
\sum_k {\rm Tr} (D_\mu\phi_k)^\dagger (D^\mu\phi_k) + \frac12 \sum_i{\rm Tr} (D_\mu \eta_i)(D^\mu\eta_i) \ ,
\ee
where 
\be
D_\mu = \partial_\mu -ig W_\mu^a T^a - \frac{i}2 g' B_\mu Y 
\ee
is the covariant derivative. In the above
$W_\mu^a$ and $g$ are the $SU(2)_L$ gauge bosons and gauge coupling, respectively, while $B_\mu$ and $g'$ are the $U(1)_Y$ gauge boson and gauge coupling. In addition, $T^a$ are the $SU(2)_L$ generators in the corresponding representation of the scalar, and $Y$ is the hypercharge generator. For complex representations we work in the basis where $T^3$ and $Y$ are diagonal. After shifting the scalar fields by their VEV's: $\phi_k\to \phi_k +\langle \phi_k \rangle$ and $\eta_i\to \eta_i + \langle \eta_i \rangle$, where the VEV's are normalized as follows
\be
\label{eq:vevnorm}
{\rm Tr}(\langle \phi_k \rangle^\dagger \langle \phi_k \rangle)=\frac12 v_k^2 \ , \qquad  {\rm Tr}(\langle \eta_i \rangle^\dagger \langle \eta_i \rangle)= \tilde{v}_i^2 \ ,
\ee
electroweak symmetry is broken and $W$ and $Z$ bosons become massive. The mass eigenstates are defined as
\bea
W^\pm &=& \frac1{\sqrt{2}}(W^1 \mp i W^2)\ , \nonumber \\  
\label{eq:eweigen}
\left( \begin{array}{cc}
             W^3\\
             B 
             \end{array}\right) &=& 
             \left( \begin{array}{cc}
             c_w & s_w \\
             -s_w & c_w
             \end{array}\right)
 \left( \begin{array}{c}
              Z\\
               A
             \end{array}\right)   ,
\eea
where the sine and cosine of the weak mixing angle are $c_w = {g}/{\sqrt{g^2+g^{\prime 2}}}$ and $s_w = {g^\prime}/{\sqrt{g^2+g^{\prime 2}}}$, respectively.  Notice the unbroken $U(1)_{em}$ leads to the conditions
\be
\left(T^3+\frac12 Y\right) \langle \phi_k \rangle= 0 \ , \qquad T^3 \langle \eta_i \rangle = 0 \ .
\ee
 Using $T^3  \langle \phi_k \rangle = -Y \langle \phi_k \rangle/2$ it is possible to express the mass terms of the $W$ and $Z$ in terms of the eigenvalues $T^2 \langle \phi_k\rangle \equiv T^aT^a  \langle \phi_k\rangle = T_k(T_k+1)  \langle \phi_k\rangle$:
\bea
\label{eq:mwgen}
m_W^2&=&\frac18 g^2\sum_{k} \left[4T_k(T_k+1) - Y_k^2\right] v_k^2 +  \frac1{2} g^2\sum_{i} T_i(T_i+1) \tilde{v}_i^2 \ , \\
\label{eq:mzgen}
m_Z^2 &=&\frac14\, \frac{g^2}{c_w^2} \sum_k Y_k^2 v_k^2 \ ,
\eea
where $Y_k$ and $Y_i$ are the hypercharges of $\phi_k$ and $\eta_i$. Couplings of the real component of the neutral scalar with the $W$ and $Z$ can be read off by the replacement $v_k \to v_k(1+\phi_k^0/v_k)$ and $\tilde{v}_i \to \tilde{v}_i(1+ \eta_i^0/\tilde{v}_i)$ in the mass terms:
\be
\Gamma^{\mu\nu}_{SV_1V_2}= g_{SV_1V_2}\,  g^{\mu\nu} \ ,
\ee
where\footnote{We include a factor of 2! when there are two identical particles in the vertex.}
\be
\label{eq:genwwzzcoup}
\begin{array}{ll}
\displaystyle g_{\phi_k WW} = \frac14 g^2 \left[4T_k(T_k+1) - Y_k^2\right] v_k \ ,  \phantom{ccc} &
\displaystyle g_{\phi_k ZZ} =  \frac12\, \frac{g^2}{c_w^2}  Y_k^2 v_k \ ,  \\
\displaystyle g_{\eta_i WW} =    g^2  T_i(T_i+1) \tilde{v}_i \ , &
\displaystyle g_{\eta_i ZZ} = 0 \ .
\end{array}
\ee
Notice that a scalar in a real representation only couples to $WW$ but not $ZZ$. Moreover,
at this order there is no scalar coupling with $Z\gamma$ and $\gamma\gamma$, which are only induced at the loop level.

At this point it is worth discussing a few examples of the $SU(2)_L$ representations appearing in the literature. The benchmark is of course the doublet Higgs scalar $H$ with $(T,Y)=(1/2,1)$. Couplings of the $CP$-even neutral Higgs $h$ with two electroweak bosons are
\be 
\label{eq:hwwzz}
g_{hWW} = \frac12 g^2 v_h \ , \quad g_{hZZ}=  \frac12\, \frac{g^2}{c_w^2}  v_h \ , \quad g_{hZ\gamma}= g_{h\gamma\gamma} = 0 \ .
\ee
Two more popular examples are the real triplet scalar $\phi$ and the complex triplet scalar $\Phi$ with $(T,Y)=(1,0)$ and  $(T,Y)=(1,2)$, respectively, for which the couplings are
\bea
\label{eq:phiwwzz}
&& g_{\phi^0WW} = 2 g^2 v_\phi \ , \quad g_{\phi^0ZZ}=  g_{\phi^0Z\gamma}= g_{\phi^0\gamma\gamma} = 0 \ , \\
\label{eq:Phiwwzz}
&& g_{\Phi^0WW} =  g^2 v_\Phi \ , \quad g_{\Phi^0ZZ}= 2 \frac{g^2}{c_w^2} v_\Phi \ , \quad  g_{\Phi^0Z\gamma}= g_{\Phi^0\gamma\gamma} = 0 \ .
\eea
We see that the $SV_1V_2$ couplings are distinctly different for scalars carrying different electroweak quantum numbers, which would give rise to different patterns of decay branching ratios into two electroweak vector bosons. However, it is well known that $\phi$ and $\Phi$ individually violate the custodial symmetry  and leads to unacceptably large corrections to the $\rho$ parameter unless the VEV is extremely small, on the order of a few GeV \cite{Amsler:2008zzb, Boughezal:2004ef, Awramik:2006uz}.

For a singlet scalar $s$, the $sV_1V_2$ couplings do not come from the Higgs mechanism. Instead, they originate from the following two dimension-five operators at the leading order:
\be
\label{eq:singletsu2}
\kappa_2 \frac{s}{4m_s} W_{\mu\nu}^a W^{a\, \mu\nu} + \kappa_1\frac{s}{4m_s} B_{\mu\nu} B^{\mu\nu}  \ ,
\ee
where the singlet $s$ is assumed to be $CP$-even.
 We have normalized the dimensionful couplings to the mass of the singlet $m_s$, although in general an unrelated mass scale could enter.
In terms of the mass eigenstate in Eq.~(\ref{eq:eweigen}), the operators become
\bea
&& \kappa_2 \frac{s}{2m_s} W_{\mu\nu}^+ W^{-\, \mu\nu} + (\kappa_2 c_w^2 + \kappa_1 s_w^2) \frac{s}{4m_s} Z_{\mu\nu}Z^{\mu\nu} \nonumber \\
&& \quad 
     + 2 c_w s_w \frac{s}{4m_s}  (\kappa_2 - \kappa_1) Z_{\mu\nu}F^{\mu\nu} + (\kappa_2 s_w^2 +\kappa_1 c_w^2) \frac{s}{4m_s}  F_{\mu\nu} F^{\mu\nu} \  .
\eea
from which we obtain the following couplings:
\bea
\label{eq:stensor}
&& \Gamma^{\mu\nu}_{sV_1V_2}= \frac{g_{sV_1V_2}}{m_s}  (p_{V_1}\cdot p_{V_2} g^{\mu\nu} -p_{V_1}^\nu p_{V_2}^\mu)  \ , \\
 \label{eq:singscoup}
&&  \begin{array}{ll}
g_{sWW}=\kappa_2 \ , \phantom{ccc} & g_{sZZ}=(\kappa_2 c_w^2 + \kappa_1 s_w^2)\ , \\
g_{sZ\gamma}= c_w s_w (\kappa_2-\kappa_1) \ , \phantom{ccc} &  g_{s\gamma\gamma} = (\kappa_2 s_w^2 + \kappa_1 c_w^2)\  .
\end{array}
\eea
One sees immediately that branching ratios following from these couplings are distinctly different from those coming from the Higgs mechanism. Moreover, the four couplings are controlled by only two unknown coefficients $\kappa_2$ and $\kappa_1$. So measurements of any two couplings would allow us to predict the remaining couplings, which, if verified experimentally, would be a striking confirmation of the singlet nature of the scalar resonance.

It is worth commenting that the coefficients $\kappa_2$ and $\kappa_1$ are related  to the one-loop beta functions of $SU(2)_L$ and $U(1)_Y$ gauge groups, respectively, via the Higgs low-energy theorem \cite{Ellis:1975ap,Shifman:1979eb}:
\begin{eqnarray}
 \beta_2(g) &=& - \frac{g^3}{(4\pi)^2} \left(\frac{11}3 C_2(G) - \frac13 n_s C(r)\right)  \ , \\
 \beta_1(g') &=& +\frac{g'^3}{(4\pi)^2} \frac13 Y^2 n_s^\prime \ .
 \end{eqnarray}
 In the above the Casmir invariants are defined as
 \be
 {\rm Tr}[t^a_r t^b_r] = C(r)\delta^{ab} \ , \qquad t_G^a t_G^b = C_2(G) \cdot \mathbf{1} \ ,
 \ee
while $n_s$ is the number of scalars in the complex representation $r$ and $n_s^\prime$ is the number of scalars charged under $U(1)_Y$. 
 Such a connection has been exploited to compute that partial width of $h\to gg$ and $h\to \gamma\gamma$ in the standard model \cite{Ellis:1975ap,Shifman:1979eb}, as well as to derive the constraints on the Higgs effective couplings \cite{Low:2009di}. For our purpose such relations serve to demonstrate that the special case of $\kappa_2=\kappa_1$, where the ratio of singlet couplings with $WW$ and $ZZ$ coincides with the standard model expectation, in general requires a conspiracy between the two one-loop beta functions to cancel each other. In this case, however, the coupling to $\gamma\gamma$ is identical to the coupling to $ZZ$.  On the other hand, depending on whether the $SU(2)_L$ running is asymptotically free,  $\kappa_2$ and $\kappa_1$ could have either the same or opposite sign, resulting in a reduction (same sign) or enhancement (opposite sign) of the $Z\gamma$ width relative to $ZZ$ and $\gamma\gamma$ channels. It is also possible that $\kappa_2=0$, resulting in a very suppressed decay width into $WW$. We will discuss further these special cases in Sect.~\ref{sect:section4}.

\section{Implications of Custodial Invariance}
\label{sect:section3}

We have seen in the previous section that scalar couplings with two electroweak bosons are uniquely determined by the $SU(2)_L\times U(1)_Y$ quantum number of the scalar involved. For nonsinglet scalars the leading contribution to the $SV_1V_2$ couplings come from the kinetic terms via the Higgs mechanism, which in turn are related to the contribution of each scalar VEV to the masses of the $W$ and $Z$ bosons. However, the ratio of the $W$ and $Z$ masses are measured very precisely and related to the precision electroweak observable $\rho=m_W^2/(m_Z c_w)^2$, which is determined at the tree-level by the structure of the scalar sector in a model. Experimentally $\rho$ is very close to 1 at the percent level \cite{Amsler:2008zzb}, which severely constrains the electroweak quantum number of any scalar which develops a VEV. 

It has been known for a long time  that the Higgs sector in the standard model possesses an accidental global symmetry $SU(2)_L\times SU(2)_R$, in which the $SU(2)_L$ and $T_R^3$ are gauged and identified with the weak isospin and the hypercharge, respectively. After electroweak symmetry breaking the global symmetry is broken down to the diagonal $SU(2)$, which remains unbroken. The unbroken $SU(2)$ is dubbed the custodial symmetry in Ref.~\cite{Sikivie:1980hm}, where it was shown the relation $\rho=1$ is protected by the custodial symmetry $SU(2)_C$. In this section we classify scalar interactions with two electroweak vector bosons according to the $SU(2)_C$ quantum number of the scalar.\footnote{In the SM the custodial invariance is explicitly broken by fermion masses, since the up-type and down-type fermions have different masses. However, this breaking is oblique in nature and only feeds into the gauge boson masses at the loop-level. Thus we do not include this particular effect in our discussion.}

There are two possibilities for the scalar sector of a model to preserve the $SU(2)_C$ symmetry. One could find a single irreducible  representation of $SU(2)_L\times U(1)_Y$  which realizes  $\rho=1$. In this case there is only one neutral $CP$-even scalar and the $W$ and $Z$ obtain masses from a single source, the VEV of the neutral scalar $S^0$.
From Eqs.~(\ref{eq:mwgen}) and (\ref{eq:mzgen}) we see the condition to realize this possibility is
\be
(2T+1)^2-3Y^2 =1\ .
\ee
An obvious solution is the Higgs doublet $(T,Y)=(1/2,1)$, beyond which the next simplest case is $(T,Y)=(3,4)$ \cite{Tsao:1980em}. However, it is clear that, since there is only one source for the masses of $W$ and $Z$ bosons, the $SV_1V_2$ couplings are derived by replacing $m_V\to m_V(1+S^0/v)$ in the mass term, which results in
\be
\label{eq:smwwzz}
g_{S^0WW}= 2\frac{m_W^2}{v} \ , \quad g_{S^0ZZ}=2\frac{m_Z^2}{v} , \quad g_{S^0Z\gamma}=g_{S^0\gamma\gamma}=0 \ .
\ee
In other words, when there is only a single source for the mass of electroweak bosons, the custodial symmetry uniquely determines the ratio of the scalar couplings to $WW$ and $ZZ$ to be
\be
\label{eq:gswwzz}
\frac{g_{S^0WW}}{g_{S^0ZZ}} = \frac{m_W^2}{m_Z^2} =  c_w^2\ ,
\ee
regardless of the $SU(2)_L\times U(1)_Y$ quantum number of the scalar involved. In the next section we will see that Eq.~(\ref{eq:gswwzz}) predicts the ratio of  the decay branching fractions into $WW$ and $ZZ$ to be roughly two-to-one, which is the case in the SM with a Higgs doublet.

The second possibility is to consider multiple scalars all contributing to the $W$ and $Z$ masses through the Higgs mechanism in such a way that, although individually the custodial invariance is not respected, the $\rho$ parameter remains 1 due to cancellations between the multiple scalars. This would happen if the scalars sits in a complete multiplet $(\mb{M}_L, \mb{N}_R)$ of the full $SU(2)_L\times SU(2)_R$ group, where $\mb{M}$ and $\mb{N}$ are positive integers labeling the $M$-dimensional and $N$-dimensional irreducible representations of $SU(2)_L$ and $SU(2)_R$, respectively. Recall that $SU(2)_L$ is fully gauged and identified with the weak isospin, while $T_R^3$ is gauged and corresponds to the $U(1)_Y$ such that $T^3_R=Y/2$, which implies the electric charge is exactly $T_C^3$:
\be
Q = T_L^3 +\frac{Y}2 = T_L^3+T_R^3 = T_C^3 \ .
\ee
Therefore, all neutral components in the scalar multiplets have $T_C^3=0$. On the other hand, unbroken custodial symmetry requires that only $SU(2)_C$ singlets are allowed to have a VEV. In other words, the scalar representation $(\mb{M}_L,\mb{N}_R)$ must contain a state with $T_C=0$, where $T_C$ is the eigenvalue labeling the quadratic Casmir operator $T_C^aT_C^a = T_C (T_C+1) \openone$. Since $T_C$ satisfies
\be
|M-N| \le T_C \le M+N\ ,
\ee
we conclude that $\rho=1$ is possible only  when $M=N$ and the scalar must furnish the $(\mb{N}_L, \mb{N}_R)$ representation. 

The trivial representation $(\mb{1}_L, \mb{1}_R)$ is a singlet scalar under $SU(2)_L\times U(1)_Y$, which was considered in the previous section. In the following we focus on the non-trivial representations, in which the $SV_1V_2$ couplings arise from the Higgs mechanism after the electroweak symmetry breaking. We will represent a scalar $\Phi_N$ in the $(\mb{N}_L, \mb{N}_R)$ multiplet in a $N\times N$ matrix whose column vectors are $N$-plets under $SU(2)_L$. The kinetic term of $\Phi_N$ is 
\bea
\label{eq:phinkin}
&& \frac12 {\rm Tr}\left[ (D^\mu\Phi_N)^\dagger D_\mu\Phi_N\right]  \ ,\\ 
&& D_\mu\Phi_N = \partial_\mu \Phi_N + i gW_\mu^a T^a \Phi_N - i g' B_\mu \Phi_N T^3 \ ,
\eea
where $T^a$ are generators of $SU(2)$ in the $N$-plet representation.
When $\Phi_N$ develops a VEV in a custodially invariant fashion\footnote{When $N$ is an odd integer, $\Phi_N$ contains a real $SU(2)_L$ $N$-plet with zero hypercharge, whose VEV has a different normalization from that in Eq.~(\ref{eq:vevnorm}): $\tilde{v}=v/\sqrt{2}$.}
\be
\label{eq:phinvev}
\langle \Phi_N \rangle = \frac{v}{\sqrt{2}}\,\openone \ ,
\ee
electroweak symmetry breaking occurs and $\rho=1$ at the tree-level.

In general various scalars in $\Phi_N$ could mix with one another and the mass eigenstates do not necessarily have well-defined $SU(2)_L\times U(1)_Y$ quantum numbers. However, it is highly desirable that the scalar potential respects the custodial symmetry so as to be consistent with $\rho=1$, which we assume to be the case. 
Then scalars with different $SU(2)_C$ quantum numbers do not mix and all the mass eigenstates have definite $SU(2)_C$ quantum numbers, according to which
we will proceed to classify the $SV_1V_2$ interactions. The $(\mb{N}_L, \mb{N}_R)$ representation decomposes under the unbroken $SU(2)_C$ as
\be
(\mb{N}_L, \mb{N}_R) = \mb{1} \oplus \mb{3} \oplus \cdots \oplus \mb{2N-3} \oplus \mb{2N-1} \ .
\ee
Scalars in the $(4k+1)$-plet are $CP$-even and those in the $(4k+3)$-plet are $CP$-odd. We assume no $CP$-violation in the scalar sector and neglect the $CP$-odd scalar interactions. Since we are interested in interactions with two electroweak gauge bosons, it is worth recalling that $W_\mu^a$ and $B_\mu$ transform as (part of) $(\mb{3}_L, \mb{3}_R)$ under $SU(2)_L\times SU(2)_R$. Therefore the only possible $SU(2)_C$ quantum numbers of a system of two electroweak gauge bosons are a singlet, a triplet, or a 5-plet, which implies the scalar must also be in one of the above three representations in order to have a non-zero coupling with two electroweak bosons. We conclude that $CP$-even $SV_1V_2$ interactions are allowed only when the scalar is either a $SU(2)_C$ singlet or a 5-plet. This is equivalent to saying two spin-1 objects can only couple to either a spin-0 or a spin-2 object. Interactions of two electroweak bosons with scalars in higher representations of $SU(2)_C$ all vanish.

Let's define the the neutral component of a custodial $n$-plet  as $H_n^0= h_n^0 X_n^0$, where $h_n^0$ is the neutral scalar field and $X_n^0$ is a $N\times N$ diagonal matrix satisfying\footnote{Recall that neutral scalars have $T_C^3=T_L^3+T_R^3=0$ and hence belong to the diagonal entries in $\Phi_N$.} 
\be
[T^a T^a, X_n^0] = n(n+1) X_n^0 \ , \qquad \qquad [T^3, X_n^0] = 0 \ , \qquad \qquad {\rm Tr}(X_n^0 X_n^0) = 1 \ .
\ee
As emphasized already, only $h_1^0$ is allowed to develop a VEV. From  Eq.~(\ref{eq:phinvev}) we see that $\langle h_1^0\rangle = \sqrt{N/2}\,v$ and $X_1^0=\openone/\sqrt{N}$, which implies all other neutral components must be (diagonal) traceless matrices:
\be
{\rm Tr}(X_n^0 X_1^0) =  {\rm Tr} (X_n^0) = 0 \ , \qquad n \ge 2 \ .
\ee
The VEV of $h_1^0$  gives rise to the following masses from the kinetic term of $\Phi_N$ :
\bea
\label{eq:mwcus}
m_W^2&=&\frac14 g^2 v^2\, {\rm Tr}\left[T^aT^a - T^3T^3\right] = \frac1{24} g^2 v^2 N(N^2-1) \ , \\
\label{eq:mzcus}
m_Z^2&=&\frac12 \, \frac{g^2}{c_w^2} v^2\, {\rm Tr}\left[ T^3 T^3 \right] =\frac1{24} \frac{g^2}{c_w^2} v^2 N(N^2-1) \ ,
\eea
which exhibits $\rho=1$. It can be verified explicitly that Eqs.~(\ref{eq:mwcus}) and (\ref{eq:mzcus}) are consistent with Eqs.~(\ref{eq:mwgen}) and (\ref{eq:mzgen}). Interactions of $h_{n}^0$, $n=1,5$, with electroweak bosons can be obtained by setting $\Phi_N = (v/\sqrt{2})\openone + H_{n}^0$ in Eq.~(\ref{eq:phinkin}):
\bea
g_{h_{n}^0WW} &=&  \frac1{\sqrt{2}} \, g^2 v\,  {\rm Tr}\left[X_{n}^0 (T^aT^a - T^3T^3)\right]  \ ,  \\
g_{h_{n}^0ZZ} &=&  \sqrt{2}\, \frac{g^2}{c_w^2} v\, {\rm Tr}\left[X_{n}^0\, T^3 T^3 \right] \ .
\eea
For the custodial singlet, $n=1$ and $X_1^0 = \openone/\sqrt{N}$, we obtain
\bea
\label{eq:h10ww}
g_{h_{1}^0WW} &=&  \frac1{\sqrt{2N}} \, g^2 v\,  {\rm Tr}\left[ (T^aT^a - T^3T^3)\right] =  2\sqrt{\frac{2}N}\frac{m_W^2}{v}  \ ,  \\
\label{eq:h10zz}
g_{h_{1}^0ZZ} &=&  \sqrt{\frac{2}N}\, \frac{g^2}{c_w^2} v\, {\rm Tr}\left[T^3 T^3 \right] =2\sqrt{\frac{2}N} \frac{m_Z^2}{v}  \ ,
\eea
which is a demonstration of the statement that any custodial singlet (apart from the one in the trivial representation $(\mb{1}_L,\mb{1}_R)$) must have couplings to the $WW$ and $ZZ$ bosons with a fixed ratio as in Eq.~(\ref{eq:gswwzz}). On the other hand, since $X_{5}^0$ is a traceless diagonal matrix, we have
\be
{\rm Tr}[X_{5}^0 T^a T^a] \propto {\rm Tr}[X_{5}^0 \openone] = 0 \ .
\ee
Then the couplings are
\bea
\label{eq:g2kww}
g_{h_{5}^0WW} &=& - \frac1{\sqrt{2}} \, g^2 v\,  {\rm Tr}\left[X_{5}^0 T^3T^3\right]  \ ,  \\
\label{eq:g2kzz}
g_{h_{5}^0ZZ} &=&  \sqrt{2}\, \frac{g^2}{c_w^2} v\, {\rm Tr}\left[X_{5}^0 T^3 T^3 \right] \ ,
\eea
which  turn out to have a ratio
\be 
\frac{g_{h_{5}^0WW}}{g_{h_{5}^0ZZ}} = - \frac{c_w^2}2 \ 
\ee
that is different from the ratio of $c_w^2$ for the custodial singlet $h_1^0$. We emphasize that  the ratios of the couplings only depend on the $SU(2)_C$ quantum numbers, and not on  the particular $(\mb{N}_L, \mb{N}_R)$ representation.

Again we discuss a few examples. The canonical example is the familiar Higgs doublet: $(\mb{2}_L, \mb{2}_R)=\mb{1}\oplus\mb{3}$, where the complex $SU(2)_L$ doublet decomposes into a singlet and a triplet under $SU(2)_C$. The $SU(2)_C$ singlet is the neutral $CP$-even Higgs, $h$, which develops a VEV and breaks the electroweak symmetry, while the triplet contains the Goldstone bosons eaten by the $W$ and $Z$. Our general expressions in Eqs.~(\ref{eq:h10ww}) and (\ref{eq:h10zz}) are consistent with those in Eq.~(\ref{eq:gswwzz}) for $N=2$.
Another example appearing in the literature \cite{Georgi:1985nv, Chanowitz:1985ug, Gunion:1989ci} is the $(\mb{3}_L, \mb{3}_R)$ representation. Under $SU(2)_L\times U(1)_Y$ it consists of a real electroweak triplet  with $(T,Y)=(1,0)$ and a complex electroweak triplet  with  $(T,Y)=(1,2)$, whose individual couplings to two electroweak bosons were summarized in Eqs.~(\ref{eq:phiwwzz}) and (\ref{eq:Phiwwzz}). In this case, the $SU(2)_C$ quantum numbers are $(\mb{3}_L, \mb{3}_R)=\mb{1}\oplus\mb{3} \oplus \mb{5}$, which contains two $CP$-even neutral scalars in the singlet and the 5-plet and one $CP$-odd scalar in the triplet \cite{Georgi:1985nv}. Our expressions for couplings of the singlet and the 5-plet with $WW$ and $ZZ$  are consistent with those in Refs.~\cite{Georgi:1985nv, Chanowitz:1985ug, Gunion:1989ci}.\footnote{Although $\rho=1$ at the tree-level in this model, constraints from $Zb\bar{b}$ vertex require $v \sim 50$ GeV \cite{Haber:1999zh}.}

It is also possible that the scalar sector of a model has multiple neutral scalar particles. In this case only scalars within the same $SU(2)_C$ multiplet are allowed to mix in order to preserve $\rho=1$. Then the ratio of the $SV_1V_2$ couplings in the mass eigenstate depends only on the $SU(2)_C$ quantum number and not on the mixing angle at all, except when there exists an electroweak singlet scalar $s$  which couples to $V_1V_2$ through the higher dimensional operators in Eq.~(\ref{eq:singletsu2}).  In this case, it is necessary to include the loop-induced couplings of $h_1^0$ with $Z\gamma$ and $\gamma\gamma$ since they are in the same order as the $sV_1V_2$ couplings. Furthermore, there could be a higher dimensional operator of the form $s|D_\mu \Phi_N|^2$, with the coefficient $\kappa_s/m_s$, which gives rise to the coupling $sV_1^\mu V_{2\, \mu}$ in addition to those in Eq.~(\ref{eq:stensor}). Even so, there are only seven  unknown parameters: $g_{h_1^0WW}$, $g_{h_1^0Z\gamma}$, $g_{h_1^0\gamma\gamma}$, $\kappa_1$, $\kappa_2$, $\kappa_s$, and the mixing angle between $h_1^0$ and $s$,  while one could measure a total of eight branching fractions of two mass eigenstates decaying into $V_1V_2$. Therefore there are enough experimental measurements to not only solve for the seven unknowns, but also test the hypothesis of mixing between $h_1^0$ and $s$. If we observe multiple scalars whose couplings to two electroweak bosons do not follow from that of $h_1^0$ or $h_5^0$, one would be motivated to consider mixing of $h_1^0$ with an electroweak singlet scalar.

\section{Partial Widths of $S\to V_1V_2^{(*)}$}
\label{sect:section4}

In this section we compute the partial decay width of $S\to V_1 V_2^{(*)}$ using the couplings derived in the previous sections. Given that the mass of the scalar could be lighter than the $WW$ threshold, we include the case of $S\to V_1V_2^*$ when one of the vector bosons is off-shell. Although decays of an electroweak doublet scalar into two electroweak bosons  have been computed both in the on-shell \cite{Lee:1977eg} and off-shell \cite{Rizzo:1980gz, Keung:1984hn, Grau:1990uu} cases, off-shell decays of an electroweak singlet scalar into two electroweak bosons do not appear to have been considered to the best of our knowledge. In the appendix we compute the decay width of a massive spin-0 particle into two off-shell vector bosons, which serve as the basis of the discussion in what follows.

From Eq.~(\ref{eq:onshellfinal}) in the appendix decays of non-electroweak singlet scalars into $WW$ and $ZZ$ are given by
\be
\label{eq:hvvgen}
\Gamma(S\to V_1V_2) = \delta_V \frac1{128\pi} \frac{|\tilde{g}_{hV_1V_2}|^2}{x^2 m_S}  \sqrt{1-4x}\, (1-4x+12x^2)  \ ,
\ee
where $x={m_{V}^2}/{m_S^2}$, $\delta_W=2$ and $\delta_Z=1$. In the limit $x^2 \ll 1$, which is a good approximation if $m_S$ is much larger than the $ZZ$ threshold, the pattern of a scalar decaying into two electroweak vector bosons is 
\be
\label{eq:doubpatt}
\Gamma(S\to WW) : \Gamma(S\to ZZ) :\Gamma(S\to Z\gamma) : \Gamma(S\to \gamma\gamma) \ \ \approx \ 2\frac{\tilde{g}_{hWW}^2}{m_W^4} : 
  \frac{\tilde{g}_{hZZ}^2}{m_Z^4} : 0 : 0  \ .
\ee
In terms of branching fractions, normalized to the branching ratio into $WW$, we have
\bea
&& {Br}_S(ZZ/WW) =  \rho^2 c_w^4 {\tilde{g}_{hZZ}^2}/{\tilde{g}_{hWW}^2}\approx  c_w^4 {\tilde{g}_{hZZ}^2}/{\tilde{g}_{hWW}^2}\ , \\
&&  {Br}_S(Z\gamma/WW) \approx {Br}_S(\gamma\gamma/WW) \approx 0 \ ,
\eea
where $Br_S(V_1V_2/WW)\equiv Br(S\to V_1V_2)/Br(S\to WW)$. Custodial symmetry then predicts unique patterns of decay branching fractions for $h_1^0$ and $h_5^0$:
\bea
\label{eq:doubpatt1}
&& {Br}_{h_1^0}(ZZ/WW) \approx \frac12 \ , \qquad {Br}_{h_1^0}(Z\gamma/WW) \approx  {Br}_{h_1^0}(\gamma\gamma/WW) \approx 0 \ ,  \\
&& {Br}_{h_5^0}(ZZ/WW) \approx 2 \ , \qquad {Br}_{h_5^0}(Z\gamma/WW) \approx {Br}_{h_5^0}(\gamma\gamma/WW) \approx 0 \ .
\eea
We see that a simple counting experiment would allow us to infer the $SU(2)_C$ quantum number of the decaying scalar!

%%%%%%%%%%%%%%%%%%%%%%%%%%%%%%%
\begin{figure}[t]
\includegraphics[scale=1]{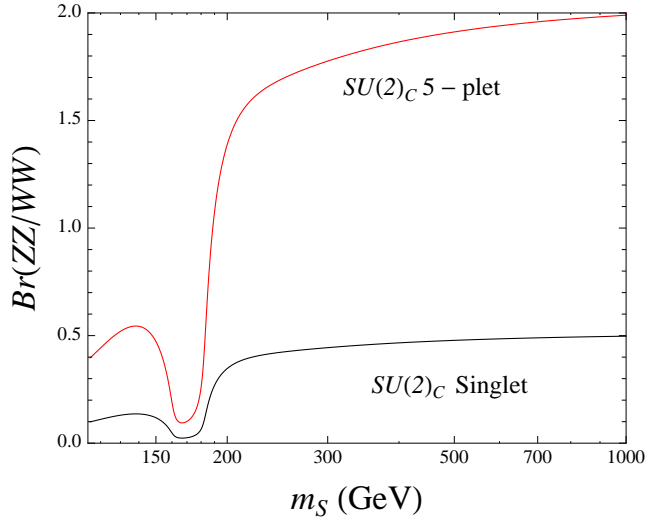}
\caption{\label{fig:fig1}\it Ratio of branching fractions into $WW$ and $ZZ$, $Br(ZZ/WW)$, for an $SU(2)_C$ singlet and a 5-plet, as a function of the scalar mass.}
\end{figure}  
%%%%%%%%%%%%%%%%%%%%%%%%%%%%%%%

In Fig.~\ref{fig:fig1} we plot the ratio $Br(ZZ/WW)$ for an $SU(2)_C$ singlet and a 5-plet, including the full kinematic dependence of the gauge boson masses, for the scalar mass between 115 GeV and 1 TeV. We include the decay into off-shell vector bosons using the expression in Eq.~(\ref{eq:offtotalwidth}) for the scalar mass below the $W$ and/or $Z$ threshold. Fig.~\ref{fig:fig1} is the unique prediction of custodial symmetry. Any deviation would imply either the electroweak singlet nature of the scalar or significant violation of custodial symmetry, which in turns suggest cancellation in the $\rho$ parameter at the percent level.

%%%%%%%%%%%%%%%%%%%%%%%%%%%%%%%%%
\begin{figure}[t]
\includegraphics[scale=1]{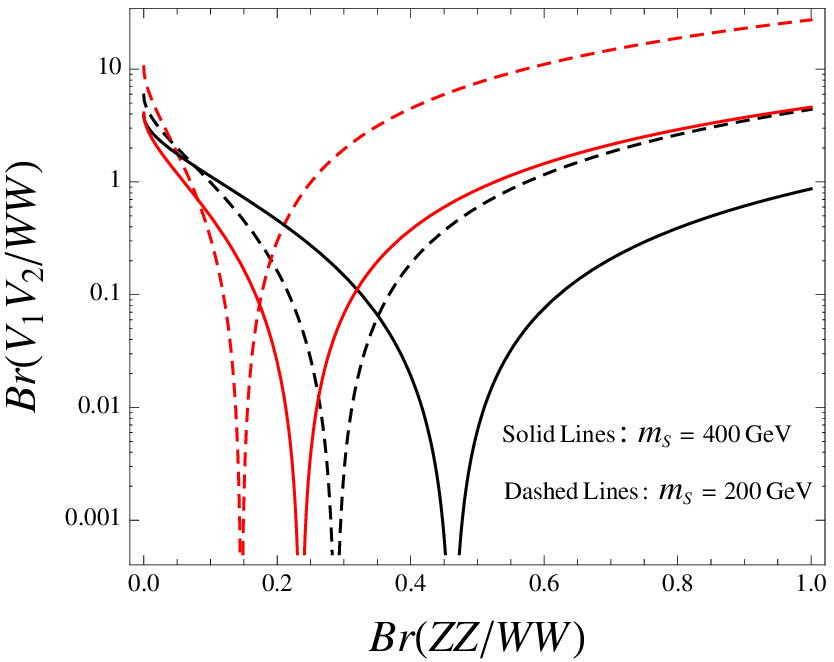}
\includegraphics[scale=1]{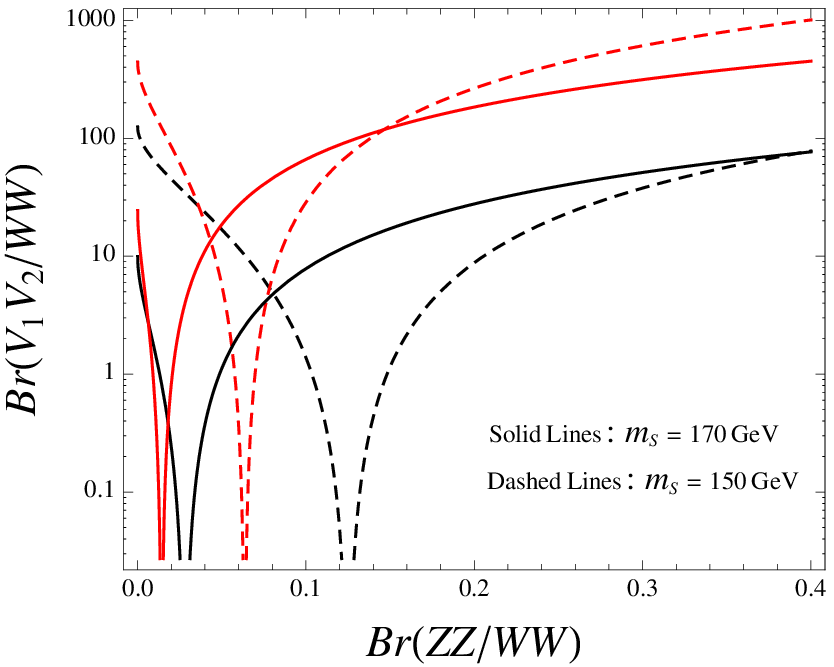}
\caption{\label{fig:fig2}\it The predicted ratios of branchings, as a function of $Br(ZZ/WW)$, for an electroweak singlet scalar. The red (gray) curves are for
$Br(\gamma\gamma/WW)$ and black curves for $Br(Z\gamma/WW)$. In this plot we assume the branching into $WW$ is nonzero.}
\end{figure}  
%%%%%%%%%%%%%%%%%%%%%%%%%%%%%%%%%%

On the other hand, using Eqs.~(\ref{eq:onshellfinal}), (\ref{eqn:onshellZgamma}), and (\ref{eq:sgaga}) in the appendix, an electroweak singlet has the following the partial decay widths into two on-shell electroweak bosons 
\bea
\label{eq:sWWwid}
\Gamma(s\to WW) &=& \frac1{32\pi} g_{sWW}^2\ m_s \sqrt{1-4x}(1-4x+6x^2) \ ,  \\
\label{eq:sZZwid}
\Gamma(s\to ZZ) &=& \frac1{64\pi} g_{sZZ}^2\ m_s \sqrt{1-4x}(1-4x+6x^2)\ , \\
\label{eq:sggwid}
\Gamma(s\to Z\gamma) &=& \frac1{32\pi} g_{sZ\gamma}^2\ m_s (1-x^2)^3 \ ,    \\
\Gamma(s\to \gamma\gamma) &=& \frac1{64\pi} g_{s\gamma\gamma}^2\, m_s \ ,
\eea
where the $g_{sV_1V_2}$ couplings are given in Eq.~(\ref{eq:singscoup}).
The pattern of partial decay widths into two electroweak bosons is then, again ignoring the effect of gauge boson masses,
\be
Br_s(V_1V_2/WW) = \delta_{V_1V_2} \frac{g_{sV_1V_2}^2}{2g_{sWW}^2}\ .
\ee
where $V_1V_2 = \{ZZ,Z\gamma,\gamma\gamma\}$, and $\delta_{V_1V_2}$ is 2 for $Z\gamma$ and 1 otherwise. This pattern is generically different from that in Eq.~(\ref{eq:doubpatt}), where the couplings arise from the Higgs mechanism. More importantly, there are only two unknowns $\kappa_1$ and $\kappa_2$. So the branching fractions into $Z\gamma$ and $\gamma\gamma$, normalized to $WW$ mode,  could be predicted as follows:
\bea
Br_s(Z\gamma/WW)&\approx&  \frac{c_w^2}{s_w^2} \left[\sqrt{2 Br_s(ZZ/WW)} -1 \right]^2 \ ,\\
Br_s(\gamma\gamma/WW)&\approx&\frac12 \left[ \frac{c_w^2}{s_w^2} \sqrt{2 Br_s(ZZ/WW)} + 1-  \frac{c_w^2}{s_w^2} \right]^2 \ .
\eea
In Fig.~\ref{fig:fig2} we plot the predicted $Br(Z\gamma/WW)$ and $Br(\gamma\gamma/WW)$ branching fractions in terms of $Br(ZZ/WW)$. Experimental verification of these relations would be a striking confirmation of the singlet nature of the scalar resonance.

By inspection of Eq.~(\ref{eq:singscoup})  we see that a special case occurs when $\kappa_2=\kappa_1$, giving
$Br_s(ZZ/WW)=1/2$, similar to that of $h_1^0$. However, in this case we have
\be
Br_s(Z\gamma/WW) \approx 0 \ , \qquad Br_s(\gamma\gamma/WW) \approx  \frac12 \ ,
\ee
up to corrections due to the mass of the $Z$ boson. By considering all four 
partial widths into the electroweak bosons it is still possible to distinguish 
a singlet scalar from the Higgs doublet even in this special case. However, 
as commented in the end of Section \ref{sect:section2}, such a scenario lacks
any obvious physical motivation.

Another special case is when $\kappa_1$=0, which occurs in the event that 
the new fermions inducing the dimension-five operators in Eq.~(\ref{eq:singletsu2}) 
carry only hypercharge and no isospin. This case is not included in Fig.~\ref{fig:fig2}
since the partial width of the scalar decaying into $WW$ vanishes! Nevertheless, 
there would still  be significant decay branching fractions into $ZZ$, $Z\gamma$, 
and $\gamma\gamma$ states, as predicted by Eq.~(\ref{eq:singscoup}).

%%%%%%%%%%%%%%%%%%%%%%%%
%Table 1 Here
%%%%%%%%%%%%%%%%%%%%%%%%%
\begin{table}[t]
\begin{tabular}{|c|c|c|c|} \hline
\makebox[3cm]{$m_S$ (GeV)} & \makebox[4cm]{$Br(ZZ/WW)$} & \makebox[4cm]{$Br(Z\gamma/WW)$} & \makebox[4cm]{$Br(\gamma\gamma/WW)$}
\\ \hline \hline
130 & 0.13 (0.13) & $4.3\times 10^{-2}$  ($6.7\times 10^{-3}$) & $3.8\times 10^{2}$ ($7.8\times 10^{-3}$)  \\ \hline
150 & 0.12 (0.12) & $1.9\times 10^{-2}$ ($3.5\times 10^{-3}$) &  65 ($2.0\times 10^{-3}$)\\ \hline
170 & $2.3 \times 10^{-2}$ ($2.3 \times 10^{-2}$) & $7.8\times 10^{-2}$ ($4.1\times 10^{-4}$) & 1.9 ($1.6\times 10^{-4}$) \\ \hline
200 & 0.36 (0.36) & $7.3\times 10^{-2}$ ($2.4\times 10^{-4}$) & 3.3 ($\alt 10^{-4}$) \\ \hline
300 & 0.44 (0.44) & $1.1\times 10^{-3}$ ($\alt 10^{-4}$) & $0.91$  ($\alt 10^{-4}$) \\ \hline
400 & 0.47 (0.47) &  $\alt 10^{-4}$ ($\alt 10^{-4}$) & $0.68$  ($\alt 10^{-4}$)  \\ \hline
\end{tabular}
\caption{\label{table1}\em Ratios of branching fractions for an electroweak singlet scalar when $Br(ZZ/WW)$ is tuned to the SM value. The value in the parenthesis is for the corresponding SM prediction.}
\end{table}
%%%%%%%%%%%%%%%%%%%%%%%%%%%

In Table I we list the ratios of branching fractions for an electroweak singlet, when $Br_s(ZZ/WW)$ of the scalar is ``tuned'' to fake that
of a SM Higgs doublet. We see in all cases $Br_s(Z\gamma/WW)$ and  $Br_s(\gamma\gamma/WW)$ are enhanced over that of the
SM ratios, especially in the low mass region, when the difference could reach five orders of magnitude at $m_S=130$ GeV for  $Br_s(\gamma\gamma/WW)$. The reason behind the enhancement is quite easy to understand: the singlet coupling strengths to all four vector boson pairs are all in the same order. Thus decays into massive final states such as $ZZ$ and $WW$ are suppressed due to phase space and kinematic factors, especially in the low scalar mass region when $WW$ and $ZZ$ channels are off-shell.
 To the contrary, in the SM the Higgs couplings to $WW$ and $ZZ$ arise at the tree-level while the couplings to $Z\gamma$ and $\gamma\gamma$  come from dimension-five operators at the one-loop level. So decays into massive final states could still dominate even below the kinematic threshold.

Another interesting case is exhibited in Table II, where $Br_s(\gamma\gamma/WW)$ is dialed to fake that of the SM Higgs. In this case the $ZZ$ channel is  suppressed relative to the $WW$ channel, while the $Z\gamma$ channel is significantly enhanced. The importance of $Z\gamma$ decays is notable, since this channel is so far neglected in the physics planning of the LHC experiments.

%%%%%%%%%%%%%%%%%%%%%%%%%%%
%Table 2 Here
%%%%%%%%%%%%%%%%%%%%%%%%%%%%
\begin{table}[t]
\begin{tabular}{|c|c|c|c|} \hline
\makebox[3cm]{$m_S$ (GeV)} & \makebox[4cm]{$Br(\gamma\gamma/WW)$} & \makebox[4cm]{$Br(ZZ/WW)$} & \makebox[4cm]{$Br(Z\gamma/WW)$}
\\ \hline \hline
115 & $2.7\times 10^{-2}$ ($2.7\times 10^{-2}$) & $5.1\times10^{-2}$ (0.11) & 39 ($9.0\times 10^{-3}$)  \\ \hline
120 & $1.7\times 10^{-2}$ ($1.7\times 10^{-2}$) & $5.7\times 10^{-2}$ (0.11) & 35 ($8.2\times 10^{-3}$)\\ \hline
130 & $7.8\times 10^{-3}$ ($7.8\times 10^{-3}$) & $6.7\times 10^{-2}$ (0.13) & 26 ($6.7\times 10^{-3}$) \\ \hline
140 & $4.0\times 10^{-3}$ ($4.0\times 10^{-3}$) & $7.1\times 10^{-2}$ (0.14) & 18 ($5.1\times 10^{-3} $) \\ \hline
150 & $2.0\times 10^{-3}$ ($2.0\times 10^{-3}$) & $6.4\times10^{-2}$ (0.12) & 10 ($3.5\times 10^{-3}$) \\ \hline
170 & $1.6\times 10^{-4}$ ($1.6\times 10^{-4}$)&  $1.4 \times10^{-2}$ ($2.3 \times 10^{-2}$) & $0.81$ ($4.1\times 10^{-4}$)  \\ \hline
\end{tabular}
\caption{\label{table2}\em  Ratios of branching fractions for an electroweak singlet scalar when $Br(\gamma\gamma/WW)$ is tuned to the SM value. The value in the parenthesis is for the corresponding SM prediction.}
\end{table}
%%%%%%%%%%%%%%%%%%%%%%%%%%%%%%

If one makes the assumption that the individual partial decay width of a scalar decaying into to $V_1V_2$ could be obtained, presumably in a future lepton collider or with a very high integrated luminosity at the LHC, then we could explore the possibility of determining the $(\mb{N}_L,\mb{N}_R)$ multiplet structure under $SU(2)_L \times SU(2)_R$. The specific question one could ask, given that the $SU(2)_C$ singlet from all $(\mb{N}_L,\mb{N}_R)$ multiplet has the same ratio of couplings to $WW$ and $ZZ$,  is whether it is possible to distinguish the $SU(2)_C$ singlet contained in  a $(\mb{2}_L,\mb{2}_R)$ from that contained in a $(\mb{3}_L,\mb{3}_R)$. To this end we observe that the couplings, $g_{h_{1}^0WW}$ and $g_{h_{1}^0ZZ} $
in Eqs.~(\ref{eq:h10ww}) and (\ref{eq:h10zz}), and the gauge boson masses in Eqs.~(\ref{eq:mwcus}) and (\ref{eq:mzcus}) are given by two parameters: $N$ and the scalar VEV $v$. Solving for $v$ in terms of the masses and $N$ we obtain
\be
g_{h_1^0WW}=g_{h_1^0ZZ}\, c_w^2= \sqrt{\frac{N^2-1}3} \, g m_W \ ,
\ee
Therefore the coupling becomes stronger as $N$ increases. The Higgs doublet has $N=2$, while the coupling of the $h_1^0$ in the $(\mb{N}_L,\mb{N}_R)$ is $\sqrt{(N^2-1)/3}$ times larger than that in the Higgs doublet, resulting in a partial decay width that is $(N^2-1)/3$ enhanced. Once $N$ is known, the complete $SU(2)_L\times U(1)_Y$ quantum number of the scalar resonance is determined.

%In the event that the SM-like ($2_L$, $2_R$) doublet is excluded as the progenitor of
%neutral scalar field $S$, we can explore to what extent it may be possible to directly
%determine the ($N_L$, $N_R$) multiplet structure under $SU(2)_L \times SU(2)_R$.
%Since we have already seen that a $SU(2)_L \times SU(2)_R$ singlet implies
%distinctive relations between different $VV$ decay channels, the proximate
%question is whether one could experimentally distinguish ($3_L$,~$3_R$)
%from ($4_L$,~$4_R$), i.e. distinguish $N$$=$$3$ from $N$$=$$4$.
%Since the couplings $g_{h_{1}^0WW}$, $g_{h_{1}^0ZZ} $
%in (\ref{eq:h10ww}), (\ref{eq:h10zz}) are proportional to $1/\sqrt{N}$,
%the corresponding partial decay widths $\Gamma_{WW}$, $\Gamma_{ZZ}$ are proportional to $1/N$.
%Thus at a lepton collider, where it should be possible to measure the
%total width $\Gamma$ of the scalar resonance, a direct measurement of $N$
%seems feasible.

As an example, at the LHC one could consider the production of the scalar in  the vector boson fusion channels
$WW/ZZ \to S \to WW$ and $WW/ZZ \to S \to ZZ$, which provide estimates of
\bea
\label{eq:vbfguys}
(\Gamma_{WW} + \Gamma_{ZZ})\frac{\Gamma_{WW}}{\Gamma_t}
\;\;{\rm and}\; \;
(\Gamma_{WW} + \Gamma_{ZZ})\frac{\Gamma_{ZZ}}{\Gamma_t}
\; .
\eea
The total width $\Gamma_t$ could be extracted by measuring the Breit-Wigner shape of the invariant mass spectrum in the $ZZ$ channel. Then one could simply fit the partial widths $\Gamma_{WW}$ and $\Gamma_{ZZ}$ using the different hypothesis for $N$. Since the event rate in this case is proportional to $\Gamma_{WW/ZZ}^2$, if the total width remains the same the enhancement of a $N\ge 3$ multiplet over the Higgs doublet is $(N^2-1)^2/9\ge 64/9\approx 7$, which is a significant enhancement.

\section{Discussion and outlook}
\label{sect:section5}

We have performed a general analysis up to dimension five of the couplings
between electroweak vector boson pairs $V_1V_2$ and a Higgs look-alike $S$, assumed to
be a neutral $CP$-even scalar resonance. We used the framework of unbroken
custodial symmetry to group the possibilities into three ``pure cases'':
scalars whose electroweak properties match a SM Higgs, scalars that are
$SU(2)_L\times SU(2)_R$ singlets and thus couple to $V_1V_2$ only at dimension
five, and scalars that couple to $V_1V_2$ as a 5-plet under custodial $SU(2)_C$.

Fig.~\ref{fig:fig1} shows that it should be straightforward to experimentally
distinguish the 5-plet case from the SM-like case of a custodial singlet, using
just the ratio of the $ZZ$ and $WW$ decay rates. 
Fig.~\ref{fig:fig2}
illustrates that $SU(2)_L\times SU(2)_R$ singlets produce distinctive
relations between the various ratios of $V_1V_2$ decay rates, emphasizing the
importance of detecting all four decay channels: $WW$, $ZZ$, $\gamma\gamma$,
and $Z\gamma$. 

To implement our proposal one can either try to extract ratios of partial decay widths directly \cite{Zeppenfeld:2002ng}, or measure the individual partial decay widths into pairs of electroweak vector bosons first \cite{Duhrssen:2004cv, Lafaye:2009vr} and then take the ratios. In the first possibility the event rate measured in each decay channel of a scalar resonance $S$ is given by 
\be
B\sigma(V_1V_2)= \sigma(S)\times Br(S\to V_1V_2) \ .
\ee
Therefore one could approximate the ratio of partial decay widths by the ratio of event rates in each channel, which are measured directly in collider experiments. It would be interesting to study ways to improve on the uncertainty arising from either possibilities.

%Moreover, we expect the uncertainty in the ratio would be smaller than that in the partial width, as some of the common uncertainties should partially cancel in  the ratio.

Since experimental analyses are often driven by final states observed, our study demonstrates the importance of having a correlated understanding of all decay channels into pairs of electroweak vector bosons to avoid misidentification. Tables I and II show how one can be badly fooled by measuring only two of the
electroweak $V_1V_2$ decay channels for a candidate Higgs. The tables were generated from
the predicted properties of a neutral $CP$-even spin 0 ``Higgs'' that
is in fact an $SU(2)_L \times SU(2)_R$ singlet imposter. In Table 1 the coefficients
$\kappa_1$, $\kappa_2$ of the dimension-five operators in Eq.~(\ref{eq:singletsu2})
have been adjusted so that the ratio of branching fractions of  $S\to ZZ$ over
$S\to WW$ coincides with the SM value for the given masses $m_S$.
In Table II the same coefficients
have been adjusted so that the branching ratio of $S\to \gamma\gamma$ over
$S\to WW$ coincides with the SM value.
In both cases measurement of the two remaining $V_1V_2$ decay rates
unmasks the Higgs imposter in dramatic fashion.

%Suppose we compute the partial widths $\Gamma_{WW}$, $\Gamma_{ZZ}$ using
%(\ref{eq:h10ww}), (\ref{eq:h10zz}) under the assumption that $N$$=$$4$,
%when in fact $N$$=$$3$. If we then use the measured observables (\ref{eq:vbfguys}) 
%to extract the total width $\Gamma$, the extracted value will be too small by a
%factor of $9/16$, within errors. Suppose also that we have measured a number
%of partial decay widths in the same production channel (either gluon fusion or VBF), to get a lower bound
%on the total width of the form
%\be
%\label{eq:lowerb}
%\frac{\Gamma}{\Gamma_{WW}} \ge 1 + \frac{\Gamma_{ZZ}}{\Gamma_{WW}}
%+ \frac{\Gamma_{\gamma\gamma}}{\Gamma_{WW}} +\ldots \; ,
%\ee
%where the ratios on the right-hand side of this expression are all observables.
%Since our VBF estimate of the left-hand side of (\ref{eq:lowerb}) will be too small
%by a factor of $3/4$, one could observe a contradiction in applying this bound.
%This strategy is a simplified version of the techniques described in
%\cite{Duhrssen:2004cv} for extracting the Higgs total width at the LHC.

In a real experiment, the analysis suggested here could be folded into
hypothesis testing based on likelihood ratios designed to expose the spin and $CP$ properties of new
heavy resonances \cite{DeRujula:2010ys,Gao:2010qx}.
Higher order effects could be included, as well as the uncertainties
associated with unfolding the experimental data to extract the $S\to V_1V_2$ 
production and decay properties.

\begin {acknowledgements}
We are grateful to Marcela Carena, Riccardo Rattazzi, and Maria Spiropulu for interesting discussions, and
to Alvaro De R\'ujula for coining the phrase ``Higgs imposters''.
I.~L. was supported in part by the U.S. Department of Energy under
contracts No. DE-AC02-06CH11357 and No. DE-FG02-91ER40684.
Fermilab is operated by the Fermi
Research Alliance LLC under contract DE-AC02-07CH11359 with the
U.S. Department of Energy.
\end{acknowledgements}

\section*{Appendix}

We consider a massive spin-0 particle $S$ decaying to two off-shell vector bosons $V_1^*$, $V_2^*$. In the rest
frame of $S$, and choosing the positive $z$-axis along the direction of $V_2$, the 4-momenta can be written:
\be
p_S = (m_S,0,0,0) \, \quad p_1 =  m_1(\gamma_1, 0, 0, -\beta_1 \gamma_1) \ , \quad
p_2 = m_2(\gamma_2, 0, 0, \beta_2 \gamma_2) \ ,
\ee
where $m_1$, $m_2$ are the off-shell vector boson masses, and the boosts
factors $\gamma_1$, $\gamma_2$, $\beta_1$, $\beta_2$ are defined by
\bea
\label{eqn:cha}
\gamma_1 &=&  \frac{m_S}{2m_1}\left( 1 + \frac{m_1^2 - m_2^2}{m_S^2} \right)  \; , \quad
%\label{eqn:chb}
\gamma_2 =  \frac{m_S}{2m_2}\left( 1 - \frac{m_1^2 - m_2^2}{m_S^2} \right)  \; ,\\
\label{eqn:sha}
\beta_1\gamma_1&=& \frac{m_S}{2m_1}\sqrt{\left( 1 - \frac{(m_1+m_2)^2}{m_S^2} \right)\left( 1 - \frac{(m_1-m_2)^2}{m_S^2} \right)} \; ,\\
\label{eqn:shb}
\beta_2\gamma_2 &=&  \frac{m_S}{2m_2}\sqrt{\left( 1 - \frac{(m_1+m_2)^2}{m_S^2} \right)\left( 1 - \frac{(m_1-m_2)^2}{m_S^2} \right)} \; .
\eea
We will use the following convenient notation:
\be
\gamma_a = \gamma_1\gamma_2(1 + \beta_1\beta_2) = \ch(y_2 - y_1)\; , \qquad
\gamma_b = \gamma_1\gamma_2(\beta_1 + \beta_2) = \sh{(y_2 - y_1)} \; ,
\ee
where $y_1$ and $y_2$ are the vector boson rapidities, as well as the following useful identities:
\be
 \gamma_a^2 - \gamma_b^2 = 1 \; , \qquad 
%\label{eqn:agamident}
 \gamma_a = \frac{1}{2m_1m_2}\left[ m_S^2 - (m_1^2+m_2^2) \right] \; , \qquad
\label{eqn:bgamident}
 \gamma_b = \frac{m_S}{m_1}\beta_2\gamma_2 \; .
\ee

It is very convenient to compute the decay widths using helicity amplitudes.
For this purpose we need to choose a consistent basis for the polarization vectors
of the vector bosons:
\bea
\epsilon_2(\lambda_2=\pm) &=& \pm\frac{1}{\sqrt{2}}(0, 1, \pm i, 0) \ , \quad \epsilon_2(\lambda_2=0) =  (\beta_2\gamma_2, 0, 0, \gamma_2) \\
%\epsilon_2(\lambda_2=-1) &=& -\frac{1}{\sqrt{2}}(0, 1, -i, 0)  \ , \quad
\epsilon_1(\lambda_1=\mp) &=& \pm \frac{1}{\sqrt{2}}(0, 1,\pm i, 0)  \ , \quad
\epsilon_1(\lambda_1=0) =  (\beta_1\gamma_1, 0, 0, -\gamma_1) 
\eea
where $\lambda_1$, $\lambda_2$ label the transverse and longitudinal polarizations.

Last but not least we will also need an expression for the two-body phase space:
\bea
d\Phi_2(p_S; p_1,p_2) &=& \frac{d^3p_1 d^3p_2}{(2\pi)^3 2E_1 (2\pi)^3 2E_2} (2\pi)^4 \delta^4(p_S-p_1-p_2) \\
&=& \frac{1}{16\pi^2}\frac{\vert \vec{p}_1 \vert}{m_S} \; d{\rm cos}\,\theta \; d\phi
\eea
where $\theta$, $\phi$ are the polar and azimuthal angles between the direction of $V_2$ and
some other reference direction, e.g. the direction of the boost from the lab frame to the $S$ rest
frame, or the direction of the beam. Note that
\be
\vert \vec{p}_1 \vert = \vert \vec{p}_2\vert = m_1\beta_1\gamma_1 = m_2\beta_2\gamma_2
=\frac{m_1m_2}{m_S}\gamma_b \; .
\ee

It is important to remember that when $V_1$, $V_2$ are distinguishable particles,
we integrate $\theta$, $\phi$ over the full $4\pi$ solid angle. However when $V_1$,
$V_2$ are identical particles (e.g. two $Z$'s or two $\gamma$'s) we should only
integrate $\theta$ from zero to $\pi/2$, to avoid counting the same final state
configuration twice. Thus the angular integration gives $2\pi$ in this case, not
$4\pi$. 

The differential off-shell decay width can be written:
\be
\frac{d^2\Gamma (S\to V_1^* V_2^*)}{dm_1^2 dm_2^2}
= \frac{2\pi \delta_V}{2 m_S} \frac{m_1m_2\gamma_b}{16\pi^2m_S^2} 
\, P_1P_2 \hspace*{-10pt}
\sum_{\lambda_1,\lambda_2 = \pm,0} 
\left\vert \Gamma^{\mu\nu}_{SV_1V_2} \epsilon^*_{\mu}(\lambda_1)\epsilon^*_{\nu}(\lambda_2)
\right\vert^2
\ee
where $\delta_V =1$ for identical vector bosons and 2 otherwise. Here $\Gamma^{\mu\nu}_{SV_1V_2} $
is the $SV_1V_2$ coupling tensor that can be read off from the Lagrangian. The propagator
factors
\be
P_i = \frac{M_{V_i}\Gamma_{V_i}}{\pi}\frac{1}
{(m_i^2-M_{V_i}^2)^2+M_{V_i}^2\Gamma_{V_i}^2}
\ee
become just $\delta(m_i^2 - M_{V_i}^2)$ in the narrow width approximation. We will write the coupling tensor as
\be
\Gamma^{\mu\nu}_{SV_1V_2} = \left( \tilde{g}_{hV_1V_2} + \frac{\tilde{g}_{sV_1V_2}}{m_S} p_1\cdot p_2 \right) \; g^{\mu\nu} - \frac{\tilde{g}_{sV_1V_2}}{m_S}\, p_1^\nu p_2^\mu   \ ,
\ee
where the coupling constants $g_h$ and $g_s$ are defined as coefficients of the following operators
\be 
 \frac{\delta_V}2 \left( \tilde{g}_{hV_1V_2}\,S\, V_1^\mu V_{2\, \mu} + \frac{\tilde{g}_{sV_1V_2}}{2m_S} S\, V_1^{\mu\nu} V_{2\, \mu\nu} \right) \ .
 \ee
In the standard model $\tilde{g}_{hV_1V_2}^2=8 m_1^2 m_2^2 G_F/\sqrt{2}$ for $WW$ and $ZZ$ channels and all other couplings vanish at the tree-level, while for an electroweak singlet scalar $\tilde{g}_{hV_1V_2}=0$.  By angular momentum conservation the only nonvanishing contributions
from the helicity sums are for $(\lambda_1,\lambda_2) = (\pm,\pm)$, or $(0,0)$:
\bea
\sum_{(\lambda_1,\lambda_2)} \left| \Gamma^{\mu\nu} \epsilon_\mu^*(\lambda_1) \epsilon_\nu^*(\lambda_2)\right|^2 &=& 
\left|\tilde{g}_{hV_1V_2}\right|^2 (2+ \gamma_a^2) +\frac{m_1^2  m_2^2}{m_S^2} |\tilde{g}_{sV_1V_2}|^2 (2\gamma_a^2+1)\nonumber \\
&&   +\frac{6m_1m_2\gamma_a}{m_S}\, \Re(\tilde{g}_{hV_1V_2}\, \tilde{g}_{sV_1V_2}^*) ,
\eea
where $\Re(c)$ is the real part of the complex number $c$. Then the off-shell decay width is
\bea
\frac{d\Gamma(S\to V_1^*V_2^*)}{dm_1^2dm_2^2}& =& \frac{2\pi \delta_V}{2m_S} \frac{m_1m_2\gamma_b}{16\pi^2m_S^2}\left[ \left|\tilde{g}_{hV_1V_2}\right|^2(2+ \gamma_a^2) +
 |\tilde{g}_{sV_1V_2}|^2\,\frac{m_1^2  m_2^2}{m_S^2} (2\gamma_a^2+1) \right. \nonumber \\
 && \qquad \left.+  \Re(\tilde{g}_{hV_1V_2}\, \tilde{g}_{sV_1V_2}^*)\, \frac{6m_1m_2\gamma_a}{m_S} \right]  P_1 P_2\ .
 \eea
 The total decay width of $S\to V_1^*V_2^*$ is given by
 \be
 \label{eq:offtotalwidth}
 \Gamma(S\to V_1^*V_2^*) = \int_0^{m_S^2} dm_1^2 \int_0^{\left(m_S-\sqrt{m_1^2}\right)^2} dm_2^2\, \frac{d\Gamma(S\to V_1^*V_2^*)}{dm_1^2dm_2^2} \ .
 \ee
The above formula is valid even when the scalar mass crosses the mass thresholds of $W$ and $Z$ bosons. More explicitly, when both vector bosons are on-shell, $m_1 \to m_V$, $m_2 \to m_V$,  we have
 \bea
 \label{eq:onshellfinal}
\Gamma(S\to V_1V_2) &=& \frac{\delta_V}{32\pi m_S}\sqrt{1-4x} \left\{\left|\tilde{g}_{hV_1V_2}\right|^2 \frac1{4x^2}(1-4x+12x^2)  \right. \nonumber \\
&& \qquad \left.+  |\tilde{g}_{sV_1V_2}|^2 \frac{m_S^2}2(1-4x+6x^2) + \Re(\tilde{g}_{hV_1V_2}\, \tilde{g}_{sV_1V_2}^*)\, 3m_S(1-2x) \right \}.
\eea
For a standard model Higgs boson, $h$, we recover the well-known expression \cite{Lee:1977eg}
\be
\Gamma (h\to V_1 V_2) = \delta_V \frac{G_F}{\sqrt{2}}
\frac{m_h^3}{16\pi} \sqrt{1-4x} (1-4x+12x^2) \ .
\ee

In the case of $S \to Z^*\gamma$, we have to take into account that only the
transverse polarizations contribute, and take the limit $m_2 \to 0$.
As $m_2 \to 0$ 
\be
\frac{d\Gamma (s\to Z^* \gamma)}{dm_1^2}
= \frac{1}{32\pi} \, \vert \tilde{g}_{sZ\gamma} \vert^2\,
m_S\, \left(1-\frac{m_1^2}{m_S^2}\right)^3 \,P_1 \ .
\label{eqn:offshellZgamma}
\ee
When the $Z$ is on-shell this becomes
\be
\Gamma (S\to Z \gamma)
= \frac{1}{32\pi} \, \vert \tilde{g}_{sZ\gamma} \vert^2\,
m_S\, (1-x)^3 \ .
\label{eqn:onshellZgamma}
\ee
The width for $S\to \gamma\gamma$ follows from this
(note we divide by 2 to get the correct phase space):
\be
\label{eq:sgaga}
\Gamma (S\to \gamma\gamma)
= \frac{1}{64\pi} \, \vert \tilde{g}_{s\gamma\gamma} \vert^2\,
m_S \ .
\ee

\end{document}